\documentclass[runningheads]{llncs}

\usepackage[utf8]{inputenc}

\usepackage{amsmath,graphicx,mathabx,bbm, color}
\usepackage[normalem]{ulem}

\allowdisplaybreaks

\begin{document}
\title{Four state deterministic  cellular automaton rule
emulating random diffusion}
%
%
\author{Henryk Fuk\'s}
\authorrunning{Henryk Fuk\'s}
%
\institute{Department of Mathematics and Statistics, Brock University, St. Catharines, ON, Canada\\
\email{hfuks@brocku.ca}}
\maketitle              
\begin{abstract}
We show how to construct a deterministic nearest-neighbour
cellular automaton (CA) with four states which
emulates diffusion on a one-dimensional lattice. 
The pseudo-random numbers needed for directing random
walkers in the diffusion process are generated with the help of rule 30.
This CA produces density profiles which agree very well
with solutions of the diffusion equation, and we discuss
this agreement for two different boundary and initial conditions. We also show how our construction can be generalized to higher dimensions.

\end{abstract}
   
\section{Introduction}
Modeling of diffusion processes with cellular automata (CA) is almost as old as the field of cellular automata itself. 
Lattice gas automata models \cite{Hardy73} can simulate diffusion of
real gas \cite{Chopard1991} very realistically and they were extensively studied in the last several decades,
thus abundant literature of the subject exists,
including monographs and textbooks \cite{chopard_droz_1998,rothman_zaleski_1997,succi2001lattice,wolf2004lattice}. 
Various models of diffusion using lattice gases
were investigated in recent years, for example
\cite{Arita2018,Medenjak2017,Deutsch2017}

Lattice gas automata are relatively complicated compared to ``classical'' CA.
Even in the simplest HPP model \cite{Hardy73} there are up to four particles per lattice site and  each particle is characterized by one of the four
allowed velocity vectors. Moreover,
the update step consists of two substeps, movement of particles
in the direction of the velocity vector followed by the collisions step
when the directions of velocity vectors of some particles are changed.
In more advanced models, such as, for example, reactive lattice gas 
automata \cite{boon96,Voroney2000}, there are three substeps, namely interaction, randomization and propagation. In the randomization substep
the call to a pseudo-random generator is required for each lattice node.

In contrast to the above, in regular CA there are no velocity 
vectors attached to particles, and the update is done in a single time step
with no need of substeps. The lattice sites change their state simultaneously
at each time step according to a specified local rule which is purely
deterministic, thus there is no need to call a random number generator.

We argue that for some applications it would be advantageous to have  such a simple deterministic nearest-neighbour cellular automaton mimicking diffusion process, so that it could be
used as a building block for various ``complexity engineering'' tasks.
For example, it could be used to constructs solutions of classification
problems in which diffusive spreading of agents is required, like in recently proposed ``diffusive'' solution of density classification problem \cite{paper53}.

What we would like to discuss in this paper, therefore, is a model
of diffusion which is not based on lattice gas automata
but rather on ``classical'' cellular automata.
It is a model of an assembly of random walkers
which perform random walk on a lattice following
exclusion principle, that is, one lattice site can be occupied by only one walker at a time. 
\section{Construction of the rule}
Consider one dimensional lattice with lattice sites being either empty (state 0)
or occupied by a single particle (state 1). All particles simultaneously
and independently of each other decide whether to move to the left or to the
right, with the same probability 0.5 in either direction. We then simultaneously move every particle to the desired position if it is empty, otherwise the particle stays in the same place. If two particles want to move to the same empty spot, only one of them, randomly selected, is allowed to do so.
This process, which constitutes a single time step, is then repeated for
as many time steps as desired.

In order to describe the process more formally, let us 
denote by $s_i$ the state of the lattice site $i$, and let $X_i$ denote binary random variable attached to site $i$. All variables $X_i$ should be independent and identically distributed  such that $Pr\left(X_i=0\right)=Pr\left(X_i=1\right)=1/2$. 
We give the following interpretation to values of random variables $X_i$. If $s_i=1$, then $X_i=1$
($X_i=0$) 
means that movement of the particle from site $i$ to the right (left)  is 
allowed. If $s_i=0$, then
$X_i=1$
($X_i=0$) 
means that arrival from the right (left) of site $i$ is allowed. If movement or arrival is not allowed, the particle
does not move. With this notation, the state of the site $i$ at the next time step, denoted by $s_i^{\prime}$,
can be expressed as follows.
\begin{gather} 
s_i^{\prime}= s_{i}
-\underbrace{s_{i}X_{i} \left( 1-s_{i+1} \right)  \left( 1-X_{i+1}
 \right)}_{move \,\,to\,\,the\,\,right}
 -\underbrace{s_{i} \left( 1-X_{i} \right)  \left( 1-s_{i-1}
 \right) X_{i-1}}_{move \,\,to\,\,the\,\,left} \nonumber \\
 + \underbrace{\left( 1-s_{i} \right)  \left( 1-X_{i}
 \right) s_{i-1}X_{i-1}}_{arrive \,\,from \,\,the\,\,left} 
 + \underbrace{\left( 1-s_{i} \right) X_{i}s_{i+1}
 \left( 1-X_{i+1} \right) }_{arrive \,\,from \,\,the\,\,right} 
 \end{gather}
The above equation can be simplified,
\begin{gather}
s_i^{\prime}=
s_{i}
+X_{i}X_{i-1}s_{i}-X_{i}X_{i-1}s_{i-1}+X_{i}X_{i+1}x_{
i}-X_{i}X_{i+1}s_{i+1} \nonumber \\-X_{i}s_{i}+X_{i}s_{i+1}-X_{i-1}s_{i}+X_{i-1}s_{i-1}.  \label{sxeq}
 \end{gather}
It is also easy to verify that for periodic boundary conditions
on a lattice of length $L$,
$$
\sum_{i=0}^{L-1} s_i^{\prime}
=\sum_{i=0}^{L-1} s_i,
$$
meaning that the number of particles is conserved.

Eq. (\ref{sxeq}) represents a probabilistic cellular  automaton,
and if we had a way to simulate $X_i$ by some pseudo-random process, we could constructs a purely deterministic CA.
This can be done by using elementary rule 30   \cite{Shin2009,Wolfram1986r30},
\begin{equation}\label{rule30}
X_i^{\prime}=f_{30}(X_{i-1}, X_i, X_{i+1}),
\end{equation}
where $f_{30}$ denotes local function of rule 30,
which can be written as
\begin{equation}
f_{30}(x_0,x_1,x_2)
=(x_0+x_1+x_2 + x_1x_2) \mod 2.
\end{equation}
This means that at each site $i$ we have two binary
state variables, $s_i$ and $X_i$, evolving, respectively,
according to eqs. (\ref{sxeq}) and (\ref{rule30}).
We can combine them together by introducing  another variable,
$$y_i=2s_i+X_i,$$ 
so that we obtain CA with four states,  $y_i \in 
\{0,1,2,3\}$. This is a fully deterministic nearest neighbour CA given by
$$y_i^{\prime}=f(y_{i-1},y_i,y_{i+1}),$$
where $f: \{0,1,2,3\}^3 \rightarrow \{0,1,2,3\}$ is defined in the Table~\ref{tabruledef}.  Let us call
$\{0,1\}$ \emph{lower states} and $\{2,3\}$
\emph{upper states}. Lower states represent empty
sites, while upper states sites occupied by particles.
Of course this mean that empty cell can be in two states (0 or 1) 
and a particle can be in two states as well (2 or 3). These
``internal'' states are used only for generation of random numbers. 
 \begin{table}
 \caption{Rule table for the diffusive rule with four states.
The entries represent $(y_{i-1},y_i,y_{i+1})\rightarrow y_i^{\prime}$.}\label{tabruledef}
\begin{center}
 \begin{tabular}{|c|c|c|c|} \hline
(0,0,0) $\rightarrow$ 0  & (1,0,0) $\rightarrow$  1 &  (2,0,0) $\rightarrow$ 0    &     (3,0,0) $\rightarrow$ 3 \\
(0,0,1) $\rightarrow$ 1  & (1,0,1) $\rightarrow$  0 &  (2,0,1) $\rightarrow$ 1    &     (3,0,1) $\rightarrow$ 2 \\
(0,0,2) $\rightarrow$ 0  & (1,0,2) $\rightarrow$  1 &  (2,0,2) $\rightarrow$ 0    &     (3,0,2) $\rightarrow$ 3 \\
(0,0,3) $\rightarrow$ 1  & (1,0,3) $\rightarrow$  0 &  (2,0,3) $\rightarrow$ 1    &     (3,0,3) $\rightarrow$ 2 \\
(0,1,0) $\rightarrow$ 1  & (1,1,0) $\rightarrow$  0 &  (2,1,0) $\rightarrow$ 1    &     (3,1,0) $\rightarrow$ 0 \\
(0,1,1) $\rightarrow$ 1  & (1,1,1) $\rightarrow$  0 &  (2,1,1) $\rightarrow$ 1    &     (3,1,1) $\rightarrow$ 0 \\
(0,1,2) $\rightarrow$ 3  & (1,1,2) $\rightarrow$  2 &  (2,1,2) $\rightarrow$ 3    &     (3,1,2) $\rightarrow$ 2 \\
(0,1,3) $\rightarrow$ 1  & (1,1,3) $\rightarrow$  0 &  (2,1,3) $\rightarrow$ 1    &     (3,1,3) $\rightarrow$ 0 \\
(0,2,0) $\rightarrow$ 2  & (1,2,0) $\rightarrow$  1 &  (2,2,0) $\rightarrow$ 2    &     (3,2,0) $\rightarrow$ 3 \\
(0,2,1) $\rightarrow$ 3  & (1,2,1) $\rightarrow$  0 &  (2,2,1) $\rightarrow$ 3    &     (3,2,1) $\rightarrow$ 2 \\
(0,2,2) $\rightarrow$ 2  & (1,2,2) $\rightarrow$  1 &  (2,2,2) $\rightarrow$ 2    &     (3,2,2) $\rightarrow$ 3 \\
(0,2,3) $\rightarrow$ 3  & (1,2,3) $\rightarrow$  0 &  (2,2,3) $\rightarrow$ 3    &     (3,2,3) $\rightarrow$ 2 \\
(0,3,0) $\rightarrow$ 1  & (1,3,0) $\rightarrow$  0 &  (2,3,0) $\rightarrow$ 1    &     (3,3,0) $\rightarrow$ 0 \\
(0,3,1) $\rightarrow$ 3  & (1,3,1) $\rightarrow$  2 &  (2,3,1) $\rightarrow$ 3    &     (3,3,1) $\rightarrow$ 2 \\
(0,3,2) $\rightarrow$ 3  & (1,3,2) $\rightarrow$  2 &  (2,3,2) $\rightarrow$ 3    &     (3,3,2) $\rightarrow$ 2 \\
(0,3,3) $\rightarrow$ 3  & (1,3,3) $\rightarrow$  2 &  (2,3,3) $\rightarrow$ 3    &     (3,3,3) $\rightarrow$ 2 \\ \hline
\end{tabular} 
\end{center}
\end{table}
\begin{figure}
 (a)\includegraphics[width=5.4cm]{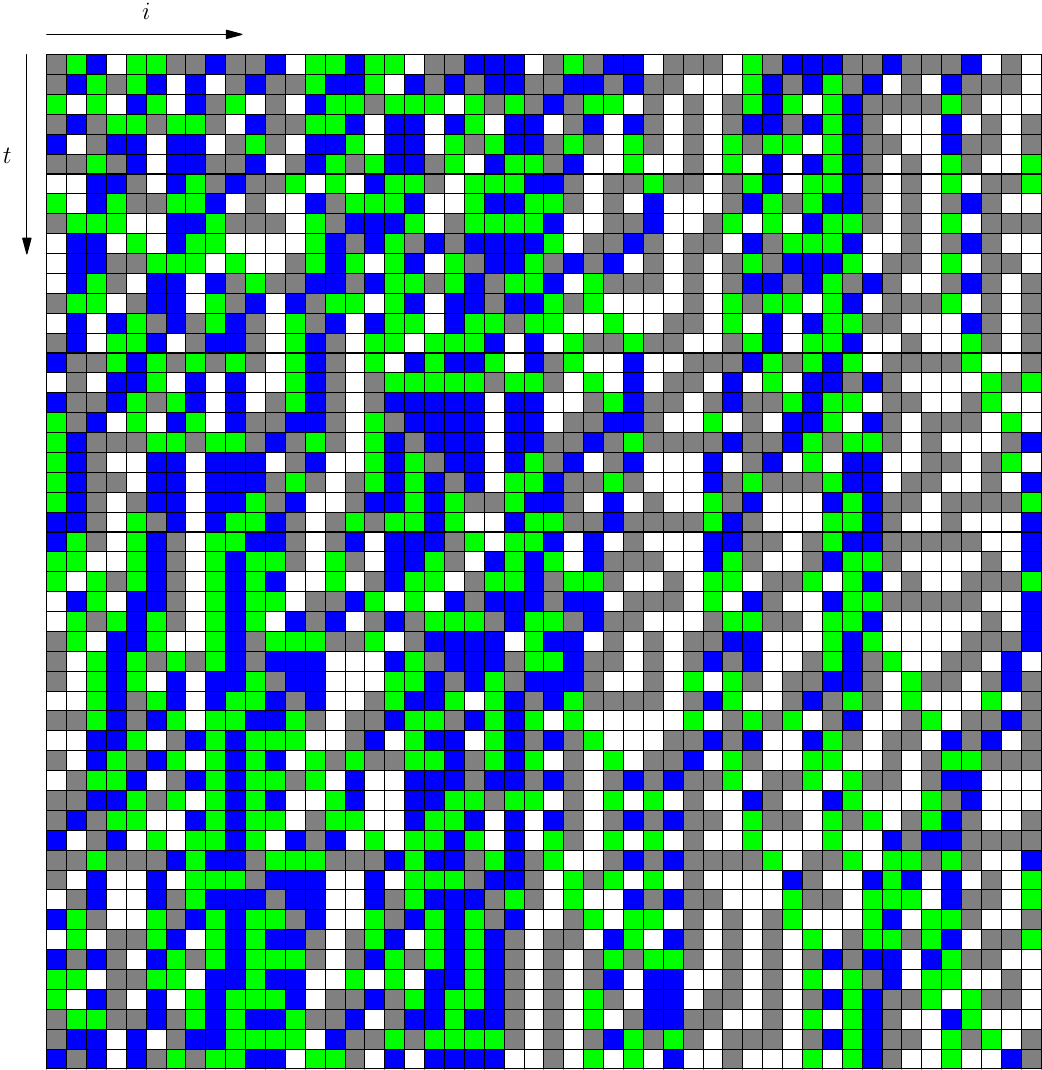}
 \includegraphics[width=5.4cm]{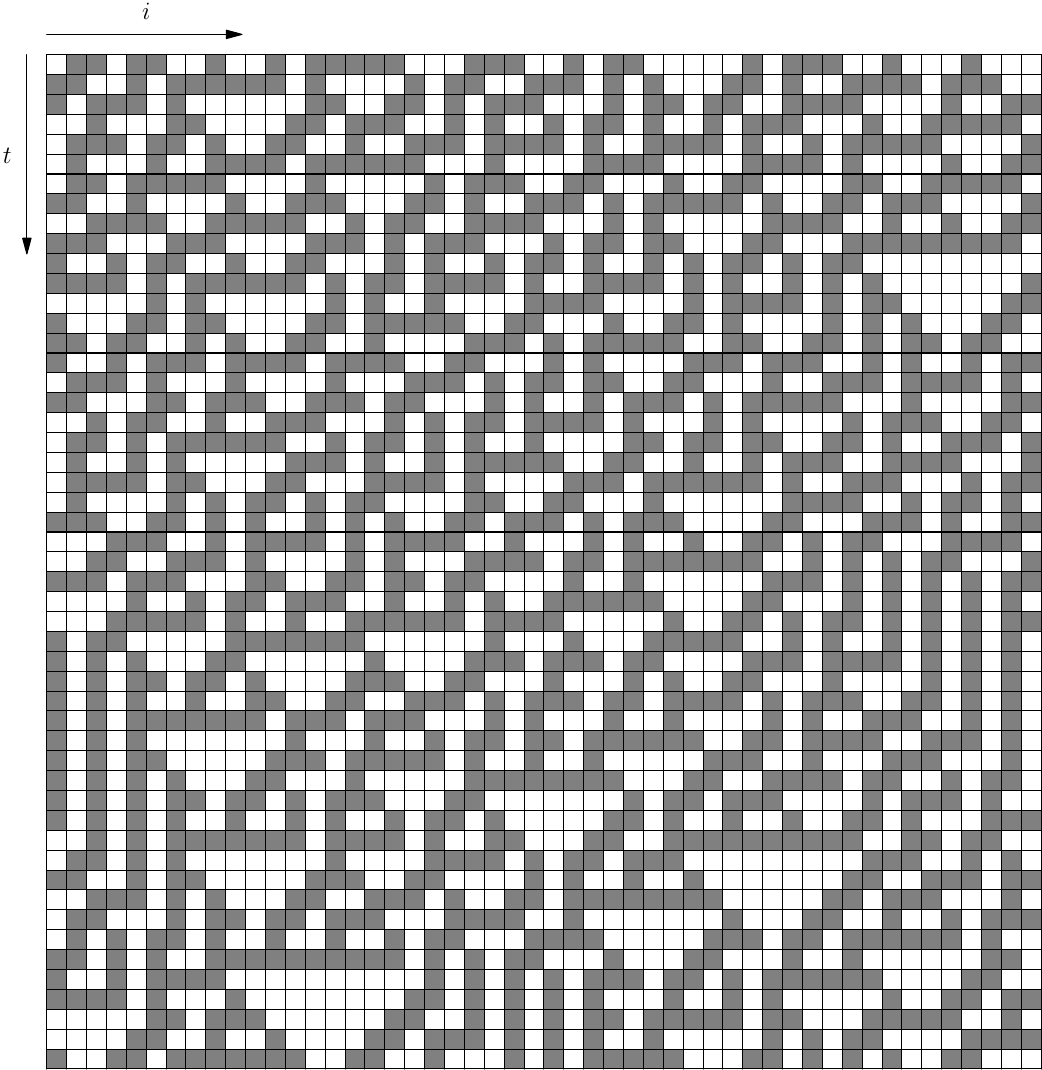}(b)\\
 (c)\includegraphics[width=5.4cm]{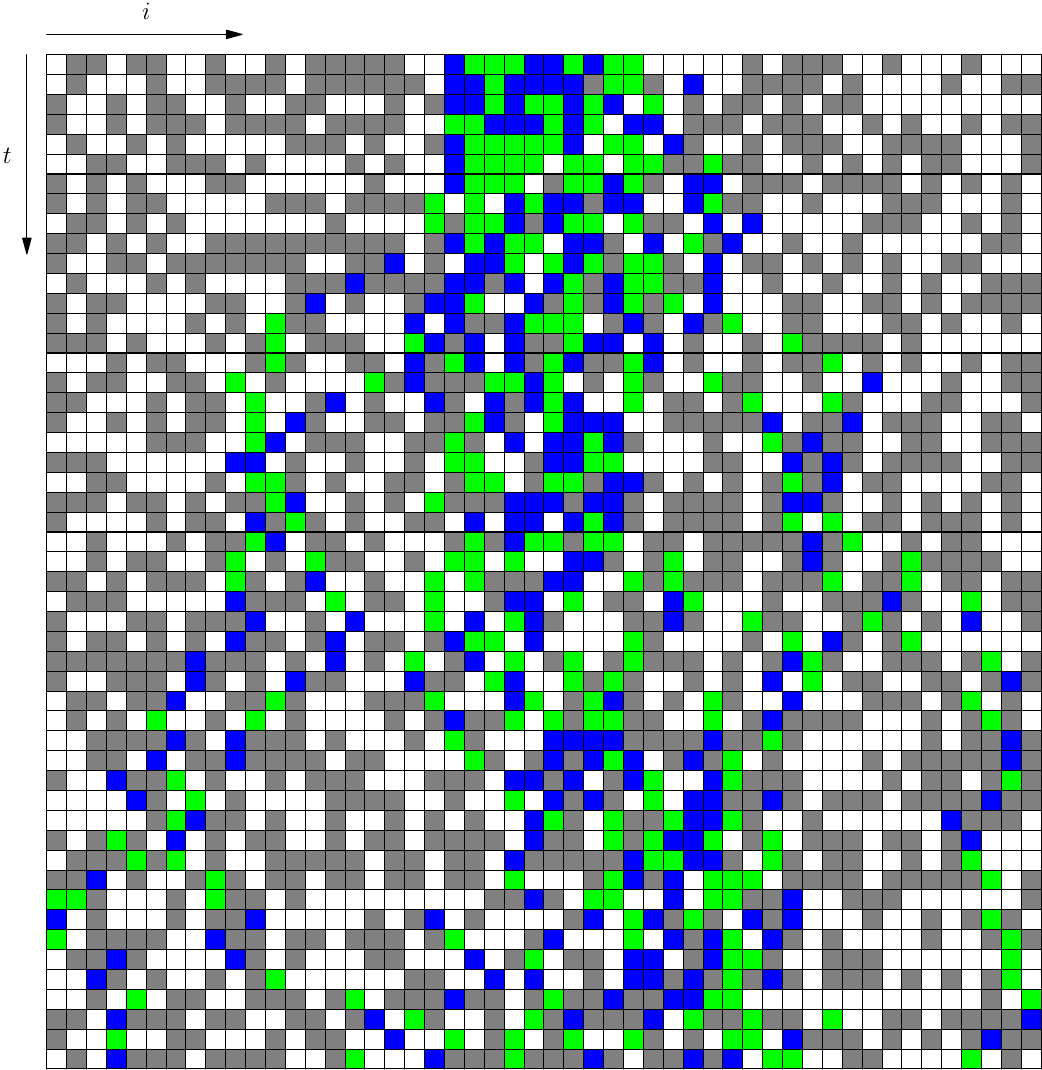}
 \includegraphics[width=5.4cm]{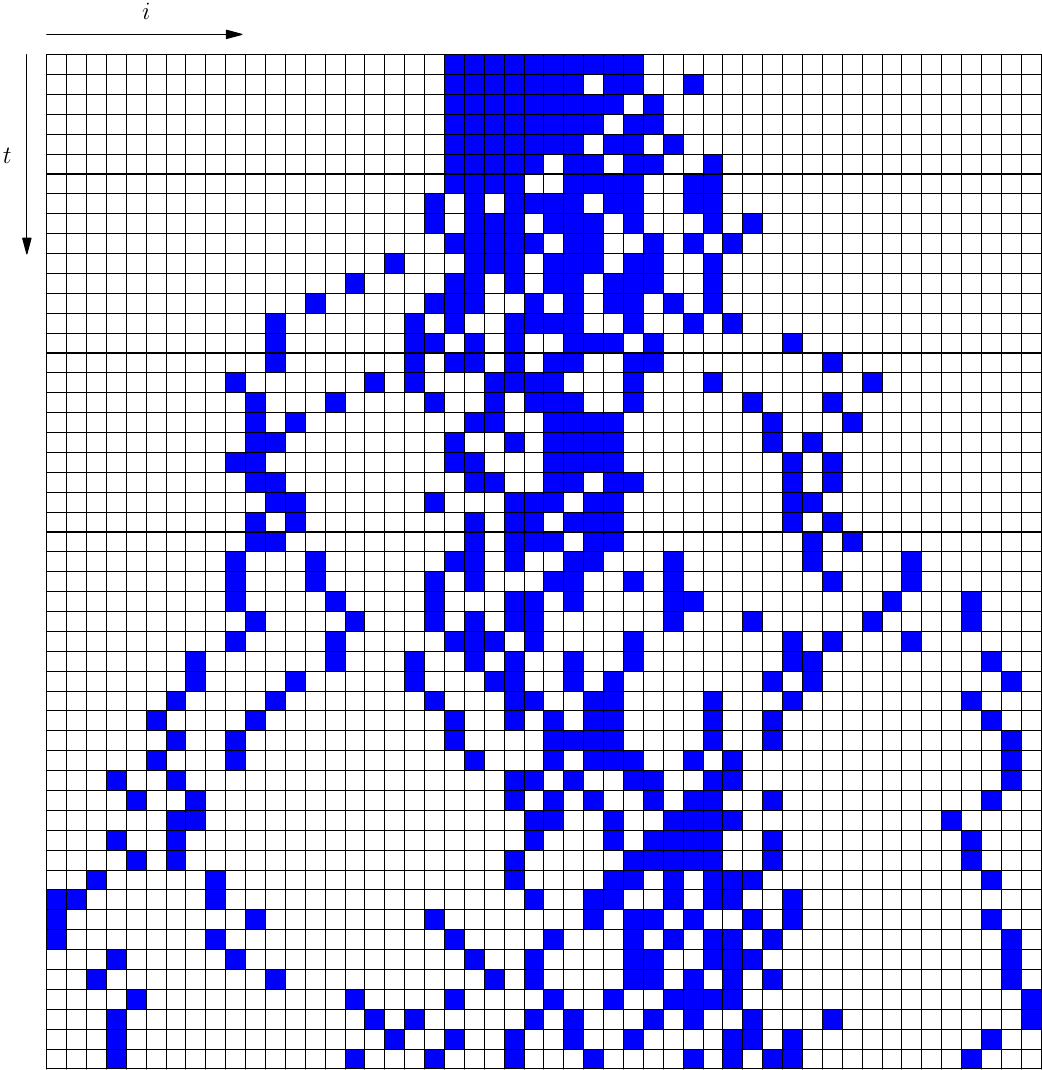}(d)
\caption{Spatiotemporal patterns for lattice with periodic
boundary conditions and length $L=50$. (a)  Random 
initial condition. (b) Random initial condition with only 0s and 1s. (c) Initial condition with block of 10 cells in upper states in the middle, lower states elsewhere; every 10th step is shown. (d) Identical pattern as in (c) but with lower states colored white and upper states blue. }
\label{fourpatterns}
\end{figure} 

Figure~\ref{fourpatterns} shows examples of spatiotemporal patterns 
produced by this rule, with upper states shown in blue/green
and lower states in grey/white. Random walk performed by 
individual particles is clearly visible. If we start with
a lattice with all sites in lower states, the well known
pattern produced by rule 30 can be observed (Fig. \ref{fourpatterns}b). 

We will now demonstrate that by taking the appropriate limit, eq.~(\ref{sxeq}) actually leads to the 
partial differential equation known as diffusion
or heat equation.
Let $\rho_i= \langle s_i \rangle$, where the angle bracket denotes the expected value.
Taking expected value of both sides of the eq. (\ref{sxeq}) we obtain
\begin{gather}
 \rho_i^{\prime}=
 \rho_i+ \frac{1}{4} \rho_i
 - \frac{1}{4} \rho_{i-1}
 + \frac{1}{4} \rho_i
 - \frac{1}{4} \rho_{i+1}
- \frac{1}{2} \rho_i
 +\frac{1}{2} \rho_{i+1}
 - \frac{1}{2} \rho_i
 + \frac{1}{2} \rho_{i-1},
 \end{gather}
where we used the fact that $\langle X_i \rangle=1/2$ for all $i$.
The above then simplifies
to
\begin{equation}
  \rho_i^{\prime}=
 \frac{1}{2}\rho_i+
 \frac{1}{4}\rho_{i+1}+
 \frac{1}{4}\rho_{i-1}.
\end{equation}
We can write this as
\begin{equation}
  \rho_i^{\prime}
  -\rho_i=
 \frac{1}{4}\left(\rho_{i+1}- 2\rho_i+
 \frac{1}{4}\rho_{i-1}\right).
\end{equation}
Let us now suppose that
the system is updated
in discrete time steps,
where the time interval between updates is $\tau$.
Moreover, let the spacing between lattice sites
be $\epsilon$. 
If we divide both sides of the above equation by $\tau$ and multiply its right hand side by 
$\frac{\epsilon^2}{\epsilon^2}$ we obtain
\begin{equation}
 \frac{ \rho_i^{\prime}
  -\rho_i}{\tau}=
 \frac{\epsilon^2}{4\tau}\frac{\rho_{i+1}- 2\rho_i+
 \rho_{i-1}}{\epsilon^2}.
\end{equation}
It is now clear that the left hand side corresponds
to numerical approximation
of the first derivative of
$\rho$ with respect to time,
while the right hand side corresponds to the  numerical approximation of the second
derivative of $\rho$ with respect to the spatial coordinate.
If we take the limit of both sides with $\epsilon
\rightarrow 0$ and, at the same time,
allowing $\tau$ tend to zero in such a way that
$\epsilon^2/\tau$ remains
constant, we get
\begin{equation}\label{heateq}
 \frac{\partial \rho}{\partial t}=
 D \frac{\partial ^2 \rho}{\partial x ^2},
 \end{equation}
where $D=\epsilon^2/4\tau$,
 $x=i\epsilon$  represents spatial coordinate, and 
 $t=k \tau$ represents time with $k$ denoting  time step, $k \in \{0,1,2,\ldots\}$. This is indeed 
 the diffusion equation. We will now show that
orbits of our rule defined in Table~\ref{tabruledef} approximate solutions
 of eq.~(\ref{heateq}) remarkably well.
\begin{figure}[t]
 (a)\includegraphics[width=5cm]{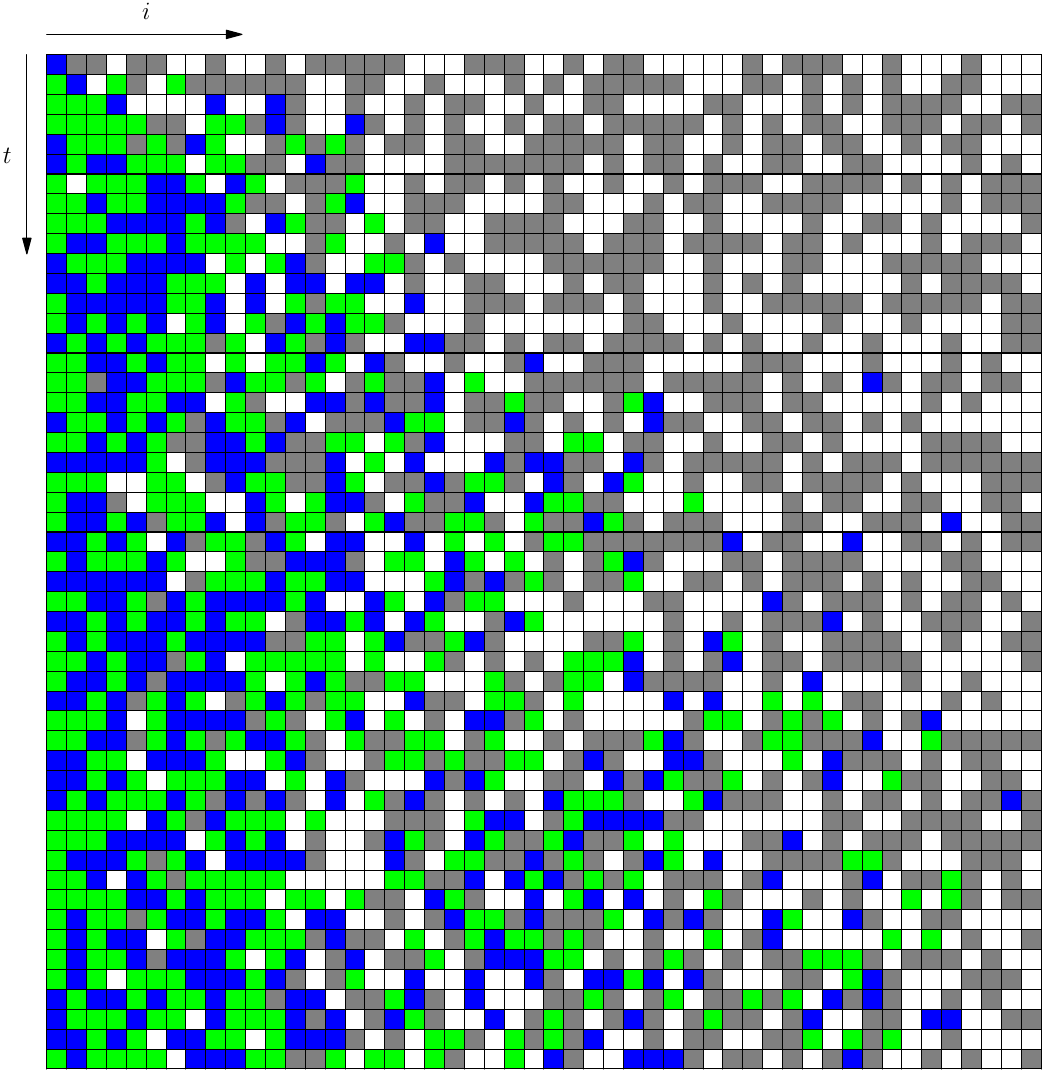}
 \includegraphics[width=5cm]{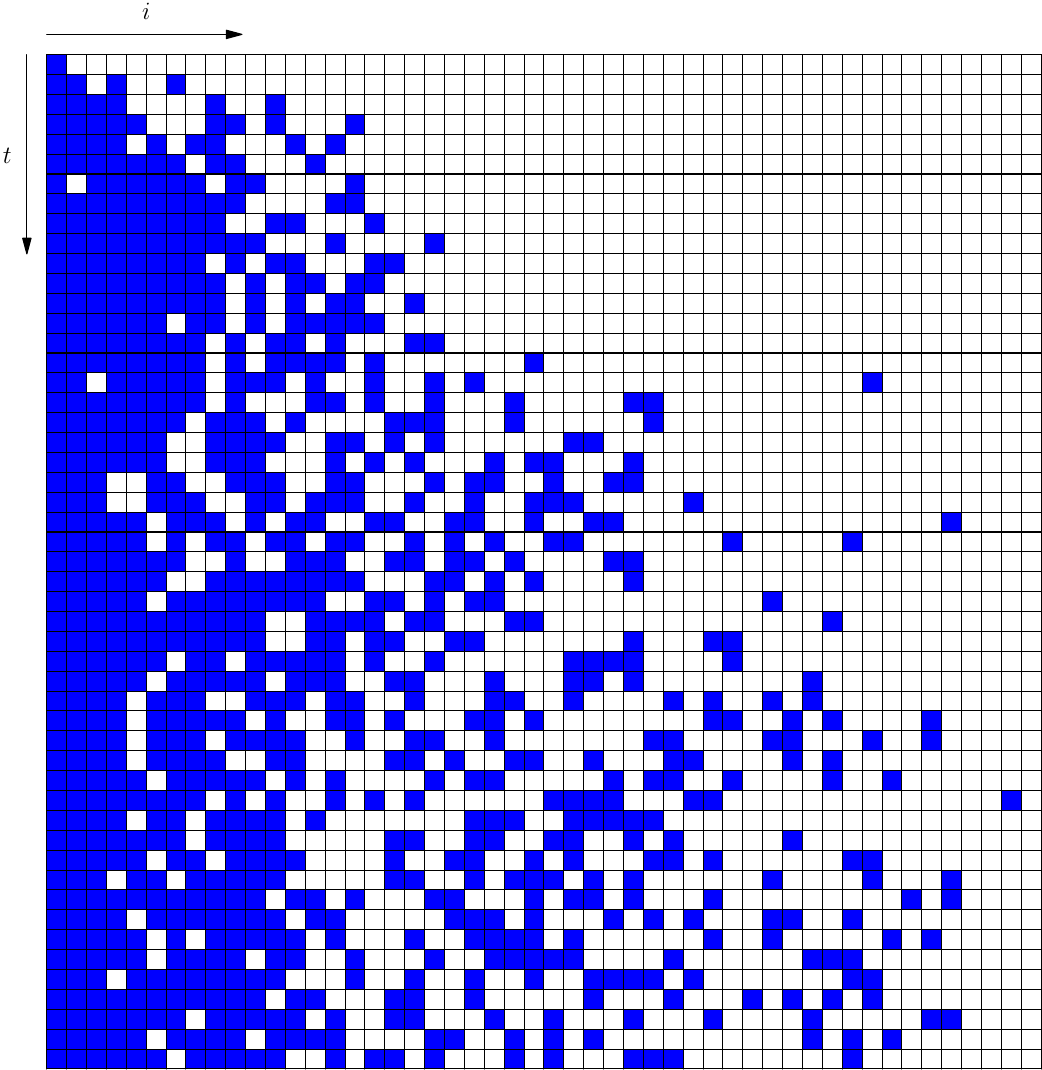}(b)
\caption{(a) Spatiotemporal patterns for lattice with fixed
boundary conditions and length $L=50$. Left end is the source of particles and the right end is kept empty. (a)  Identical pattern as in (a) but with lower states colored white and upper states  blue. Only every 10-th step is shown. }\label{inhomobar}
\end{figure} 
 \begin{figure} 
\begin{center}
 \includegraphics[width=10cm]{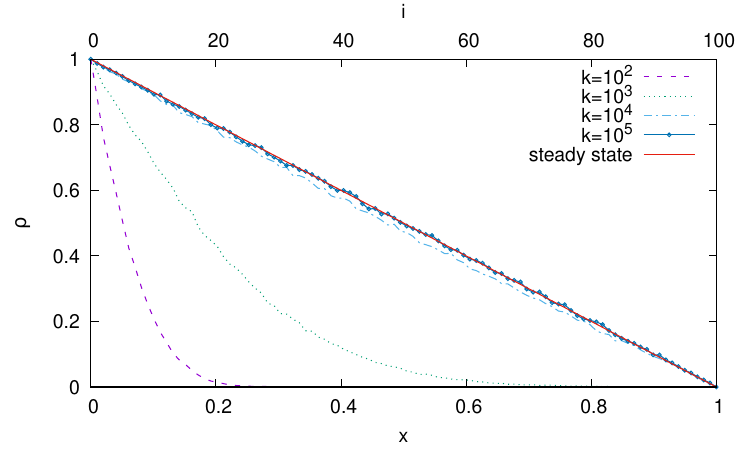}
\end{center}
\caption{Density profiles obtained by numerical experiments for finite  lattice ($L=100$) with inhomogeneous boundary conditions. Left end is the source of particles and the Right end is always empty.  Vertical axis represents density obtained after $k$ iterations, averaged over $10^4$ runs.}\label{curve1000}
\end{figure} 
\begin{figure}
\begin{center}
 (a)\includegraphics[width=12cm]{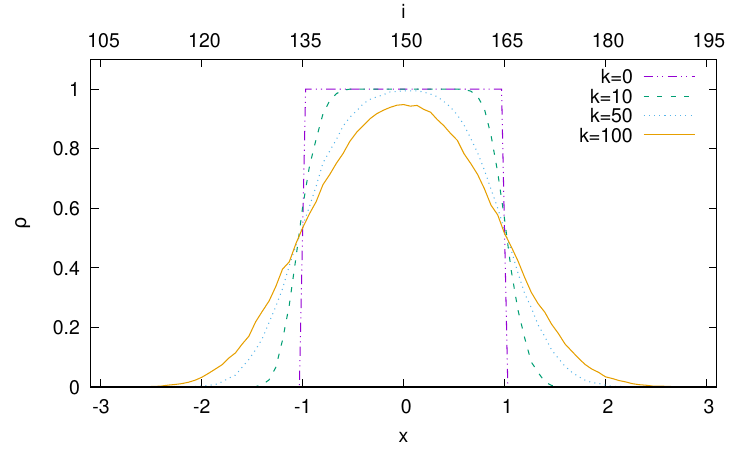}\\
  (b)\includegraphics[width=12cm]{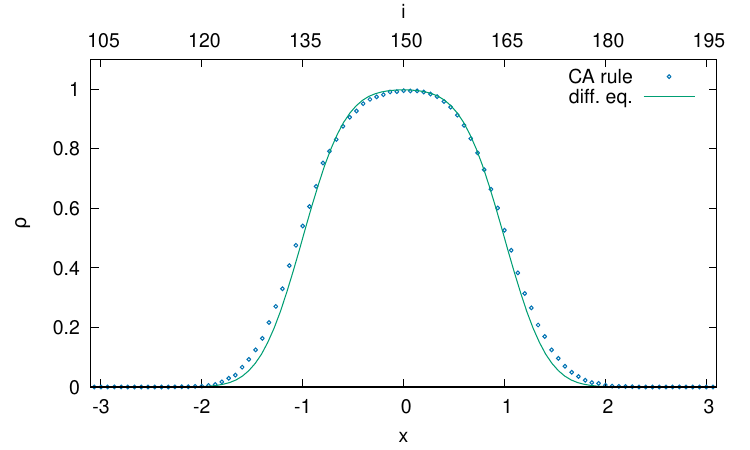}
\end{center}
\caption{(a) Development of density profile for lattice of length $L=300$ with particles
initially located only at $i=135, 136,\ldots 165$. Vertical axis represents density obtained after $k$ iterations, averaged over $10^4$ runs.
(b) Density profile for CA for $k=50$ compared 
 with solution of the diffusion equation given by eq. (\ref{solution-rect}).
}\label{developprofilen}
\end{figure} 
\section{Experiments}
We will consider two numerical experiments highlighting
the quality of the rule of Table~\ref{tabruledef}.
The first one is usually described in PDE textbooks
as a heated finite bar with inhomogeneous boundary conditions \cite{Boyce2001}. We will consider
finite lattice of size $L$ with fixed boundaries where the leftmost site is always occupied by a particle and the rightmost site is always empty. Figure~\ref{inhomobar}
shows the corresponding spatiotemporal patterns.
We computed numerical approximations of $\rho_i$
by obtaining average value of $s_i$ after $k$ iterations,
where the average is obtained by repeating the simulation
$10^4$ times. Defining $x=i/L$ we then plotted 
$\rho$ versus $x$ for various values of $k$.
Results are shown in  Figure~\ref{curve1000}.

Let us compare the results with solution of
eq~\ref{heateq} with boundary conditions
$\rho(0,t)=1$, $\rho(1,t)=0$, given by
the following~\cite{Boyce2001} infinite series,
\begin{equation}
\rho(x,t)=1-x-\frac{2}{\pi}\,\sum _{n=1}^{\infty } \frac{1}{n}  {\rm exp}\left( {-{ {{n}^{2} {\pi }^{2} D t}}} \right) \sin \left(  {n\pi x} \right).
 \end{equation}
One can see that as $t\to \infty$, corresponding to our 
$k \to \infty$, the density profile should tend to
a straight line, $\rho(x,\infty)=1-x$, labelled in Figure~\ref{curve1000} as ``steady state'' line.
For $k=10^5$ the experimental density profile 
almost overlaps with $1-x$, confirming that the 
approximation of eq.~(\ref{heateq}) by rule of Table~\ref{tabruledef} is indeed very good.
 

The second experiment we will describe is the case of
the initial configuration where all the particles
are placed in a solid block in the middle of the lattice,
just like in Figure~\ref{fourpatterns}c and \ref{fourpatterns}d. We again
computed average densities using $10^4$ runs, and the results are shown in Figure~\ref{developprofilen}a.
We used lattice of $300$ sites with only 30 sites occupied initially, for $i=135, 136, \ldots, 165$, 
the rest being empty. 
Spatial variable $i$ (upper axis) is rescaled as $x=(i-150)/15$ (lower axis), so that $x=-1$ corresponds to $i=135$ and $x=1$ corresponds  $i=165$. The rescaling was done
to compare the CA density profiles  with solution of eq.~(\ref{heateq})
with initial condition
\begin{equation*}
\rho(x,0)=\begin{cases}
          1 \quad &\text{if} \, |x|<1, \\
          0 \quad &\text{otherwise,} \\
     \end{cases}
\end{equation*}
which, following \cite{Kreyszig2011}, is given as
\begin{equation}\label{solution-rect}
 \rho(x,t)=\frac{1}{2\sqrt{D \pi t}} \int_{-1}^1 \exp \left(
 - \frac{(x-v)^2}{4 D t} 
 \right)\, dv
 ={\frac{1}{2}{\rm erf}\left({\frac {x+1}{2\sqrt {Dt}}}\right)}-
\frac{1}{2}{{\rm erf}\left({\frac {x-1}{2\sqrt {Dt}}}\right)}.
\end{equation}

For $k=50$, we compared the numerically obtained density profile (shown in Figure~\ref{developprofilen}a as dotted line)  with the corresponding solution
of the diffusion equation given by eq.~\ref{solution-rect}.
In Figure~\ref{developprofilen}b,  the density profile obtained by the CA rule for $k=50$ is 
shown together with the corresponding graph of the right hand side of 
eq.~(\ref{solution-rect}). We can again see very good agreement of both, although there are slight discrepancies in the intervals around $x=\pm 1.5$. Given that we are comparing orbits of the discrete process with solution of the
continuous PDF, the agreement is still quite remarkable.

\section{Two-dimensional rule}
\begin{figure}
\begin{center}
 \includegraphics[width=10cm]{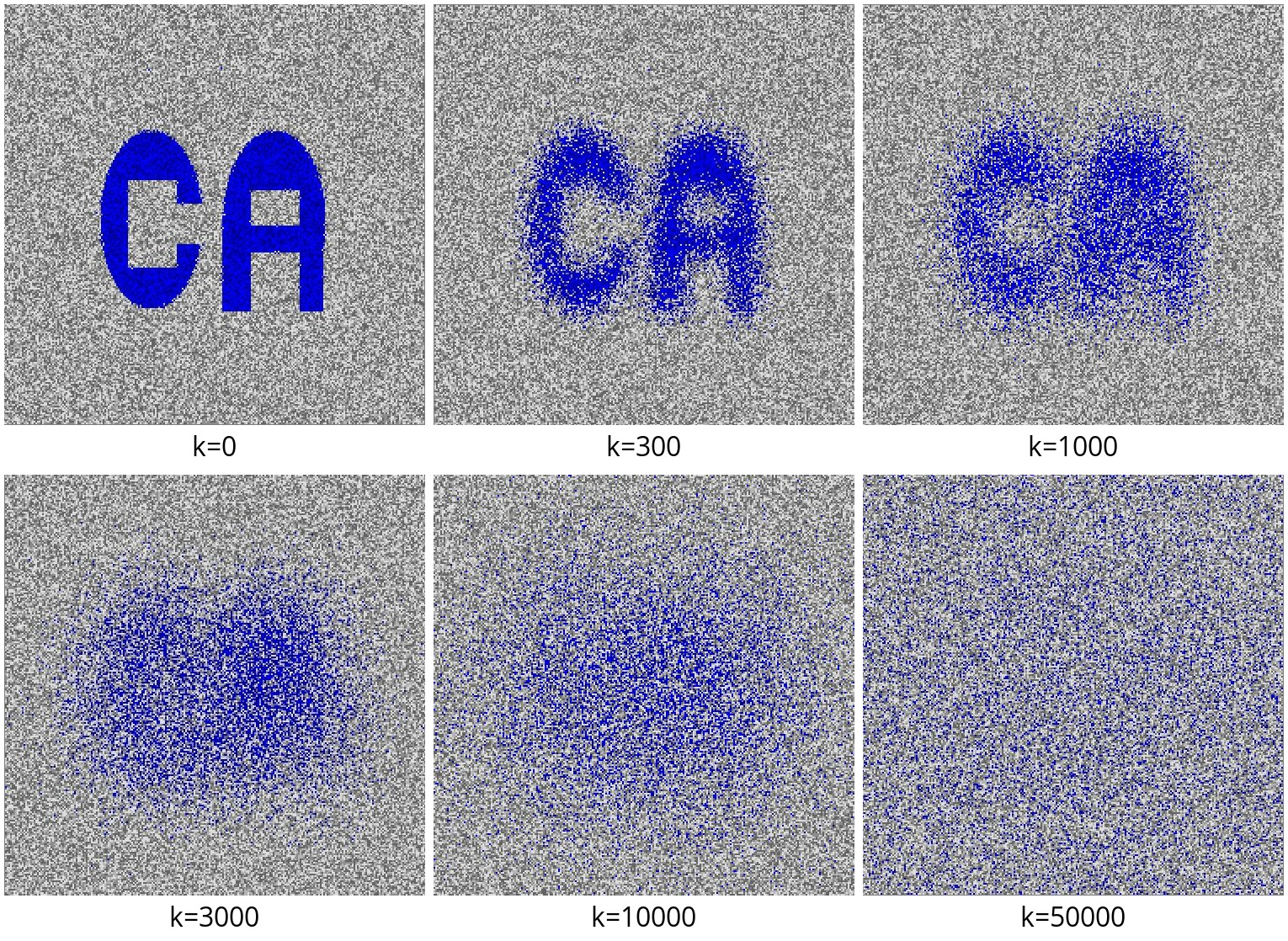}
\end{center}
\caption{Patterns produced by two-dimensional diffusive rule starting from the initial image depicting letters ``CA'' on a lattice
$250\times250$ with periodic boundary conditions.
States 0--3 are shown as shades of gray  and states 4-7 as shades of blue.}\label{2dpatterns}
\end{figure}

It is not difficult to construct the deterministic diffusion rule in higher dimensions, following
the method outlined in the first section. As an example,
we will show two-dimensional version of the rule of Table~\ref{tabruledef}. In this case, two independent pseudo-random variables $X_{i,j}$
and $Y_{i,j}$ are needed, controlling the movement in, respectively, horizontal and vertical direction. These variables can be obtained by using the rule 30 applied in horizontal and
vertical direction, 
\begin{align}
X_{i,j}^{\prime}&=f_{30}(X_{i-1,j}, X_{i,j}, X_{i+1,j}),\\ 
Y_{i,j}^{\prime}&=f_{30}(Y_{i,j-1}, Y_{i,j}, Y_{i,j+1}).
\end{align}
The two-dimensional diffusive rule is then given by
\begin{gather*}
 s_{i,j}^{\prime}=s_{i,j}
 - \underbrace{s_{i,j}       X_{i,j}      (1-s_{i+1,j})  (1-X_{i+1,j})  Y_{i,j} Y_{i+1,j}}
    _{move \,\,to\,\,the\,\,right}\\
- \underbrace{ s_{i,j}      (1-X_{i,j})  (1-s_{i-1,j})  X_{i-1,j}      Y_{i,j} Y_{i-1,j}}
     _{move \,\,to\,\,the\,\,left}\\
     + \underbrace{(1-s_{i,j})  (1-X_{i,j})  s_{i-1,j}      X_{i-1,j}      Y_{i,j} Y_{i-1,j}}
      _{arrive \,\,from\,\,the\,\,left}\\
     + \underbrace{(1-s_{i,j})  X_{i,j}        s_{i+1,j}      (1-X_{i+1,j})  Y_{i,j} Y_{i+1,j}}
     _{arrive \,\,from\,\,the\,\,right}\\
     -\underbrace{s_{i,j}       X_{i,j}      (1-s_{i,j+1})  (1-X_{i,j+1})  (1-Y_{i,j}) (1-Y_{i,j+1})}
     _{move \,\,to\,\,the\,\,top}\\
     - \underbrace{s_{i,j}      (1-X_{i,j})  (1-s_{i,j-1})  X_{i,j-1}      (1-Y_{i,j}) (1-Y_{i,j-1})}
     _{move \,\,to\,\,the\,\,bottom}\\
     + \underbrace{(1-s_{i,j})  (1-X_{i,j})  s_{i,j-1}      X_{i,j-1}      (1-Y_{i,j}) (1-Y_{i,j-1})}
     _{arrive \,\,from\,\,the\,\,bottom}\\
     + \underbrace{(1-s_{i,j})  X_{i,j}       s_{i,j+1}      (1-X_{i,j+1})  (1-Y_{i,j}) (1-Y_{i,j+1})}
      _{arrive \,\,from\,\,the\,\,top}. 
\end{gather*}
We can then introduce variable 
$$y_{i,j}=4s_{i,j}+2Y_{i,j}+X_{i,j},$$ 
and with this new variable we will obtain deterministic cellular automaton with 8 states and von Neumann neighbourhood, where lower states $\{0,1,2,3\}$ correspond to empty sites and upper states
$\{4,5,6,7\}$ to occupied sites. The rule table of this rule consists of $8^5=32768$ entries, thus it cannot be reproduced here. Nevertheless, using compression tool for CA rules
included with Golly software \cite{golly}, this
rule table can be reduced to 94 transitions using 31 variables.
The .rule file for Golly  program is available
from the author, allowing to perform interactive experiments
with the rule. Results of one of such experiments 
are shown in Figure~\ref{2dpatterns}, where we used only two colors, white for low states and blue for high states. This is done to emphasize the dynamics of the diffusion process and
to ``hide'' the generation of random variables by two
embedded rules 30.
\section{Conclusions}
Deterministic nearest-neighbour cellular automaton
modelling diffusion process with very high fidelity
can easily be constructed providing that sufficient
number of states is employed, and in $d$ dimensions
$2^{d+1}$ states are needed. This brings an interesting 
question and research challenge: could one construct realistic diffusion model
with smaller number of states? In particular, in
one dimension, can we construct a nearest-neighbour CA rule with only 3 states (instead of our 4), yet emulating
diffusion process with similar quality as the rule
presented here? The answer is most likely no,
yet one would have to formulate the problem in a more
rigorous fashion first in order to give the definitive answer. What is certain is that it cannot be done with two states, as none of the elementary CA rules  exhibits
sufficient diffusion-like properties.

\textbf{Acknowledgements:} the author acknowledges financial support from the Discovery Grant by National Science and Engineering Council of Canada.

\end{document}